\documentclass[preprint]{elsarticle}

\usepackage{lineno,hyperref}
\usepackage{amsfonts,graphicx,subfigure,bm,amssymb}
\usepackage{amsmath}
\modulolinenumbers[5]

\journal{Journal of \LaTeX\ Templates}

\begin{document}

\begin{frontmatter}

\title{Frequency Estimation of Multiple Sinusoids with Three Sub-Nyquist Channels\tnoteref{mytitlenodte}}
\tnotetext[mytitlenote]{This work is supported by the National Natural Science Foundation of China under Grant 61501335.}

\author{Shan Huang\fnref{}}
\author{Haijian Zhang\fnref{}}
\author{Hong Sun\fnref{}}
\author{Lei Yu\fnref{}}
\author{Liwen Chen\fnref{}}
\address{Signal Processing Laboratory, School of Electronic Information, Wuhan University, Wuhan 430072, China}

%


\begin{abstract}
Frequency estimation of multiple sinusoids is significant in both theory and application. In some application scenarios, only sub-Nyquist samples are available to estimate the frequencies. A conventional approach is to sample the signals at several lower rates. In this paper, we address frequency estimation of the signals in the time domain through undersampled data. We analyze the impact of undersampling and demonstrate that three sub-Nyquist channels are generally enough to estimate the frequencies provided the undersampling ratios are pairwise coprime. We deduce the condition that leads to the failure of resolving frequency ambiguity when two coprime undersampling channels are utilized. When three-channel sub-Nyquist samples are used jointly, the frequencies can be determined uniquely and the correct frequencies are estimated. Numerical experiments verify the correctness of our analysis and conclusion.
\end{abstract}

\begin{keyword}
Frequency estimation, Sub-Nyquist sampling, Coprime sampling.
\end{keyword}

\end{frontmatter}


\section{Introduction}
Frequency estimation of multiple sinusoids has wide applications in communications, audio, medical instrumentation and electric systems \cite{Funada1987A}\cite{Arablouei2015Adaptive}\cite{Besson2016Direction}. Frequency estimation methods cover classical modified discrete Fourier transform (DFT) \cite{Belega2015Frequency}\cite{candan2015fine}, subspace techniques such as "multiple signal classification" (MUSIC) \cite{schmidt1986multiple} and "estimating signal parameter via rotational invariance techniques" (ESPRIT) \cite{roy1989esprit} and other advanced spectral estimation approaches \cite{Zachariah2013Line}\cite{Tu2016Phase}. In general, the sampling rate of the signal is required to be higher than twice the highest frequency (i.e. the Nyquist rate). The sampling frequency increases as the frequencies of the signals, which results in much hardware cost in applications \cite{hill1994benefits}. In some applications, such as velocity synthetic aperture radar (VSAR) \cite{friedlander1998vsar}, the received signals may be of undersampled nature. So it is necessary to study frequency estimation from undersampled measurements. In addition, this problem has a close connection with phase unwrapping in radar signal processing and sensor networks \cite{Zhang2008Time}\cite{li2008fast}.

A number of methods have been proposed to estimate the frequencies with sub-Nyquist sampling. To avoid the frequency ambiguity, Zoltowski proposed a time delay method which requires the time delay difference of the two sampling channels less than or equal to the Nyquist sampling interval \cite{zoltowski1994real}. By introducing properly chosen delay lines, and by using sparse linear prediction, the method in \cite{tufts1995digital} provided unambiguous frequency estimates using low A/D conversion rates. The authors of \cite{li2009robust} made use of Chinese remainder theorem (CRT) to overcome the ambiguity problem, but only single frequency determination is considered. Bourdoux used the non-uniform sampling to estimate the frequency \cite{bourdoux2011sparse}. Some scholars used multi-channel sub-Nyquist sampling with different sampling rates to obtain unique signal reconstruction \cite{venkataramani2000perfect}\cite{sun2012wideband}. These methods usually impose restriction on the number of the frequency components, which depends on the number of the channels. Based on emerging compressed sensing theory, sub-Nyquist wideband sensing algorithms and corresponding hardware were designed to estimate the power spectrum of a wideband signal \cite{tropp2010beyond}\cite{mishali2010theory}\cite{Wang2016Novel}. However, these methods usually require random samples, which often leads to complicated hardware, making the practicability discounted. In \cite{Pal2011Coprime} and \cite{Vaidyanathan2011Sparse}, two channels with coprime undersampling ratios are utilized to estimate line spectra of multiple sinusoids. By considering the difference set of the coprime pair of sample spacings, virtual consecutive samples are generated from second order moments \cite{Qin2015Generalized}. The method only requires double sub-Nyquist channels without additional processing, the hardware is simpler than the most of former methods.

In this paper, we use three channels other than two channels with coprime undersampling ratios to get enough data. It is demonstrated that the estimated frequencies sometimes can not be uniquely determined when only two channels with coprime undersampling ratios are utilized. In the sampling scheme of multiple channels, if the ambiguous frequencies estimated from single channel are matched successfully, the correct estimated frequencies will be found \cite{Xia2000An}. Through the analysis for the matching process, we deduce the condition that leads to the failure of resolving frequency ambiguity. With the samples obtained from the three channels, the MUSIC algorithm is used to estimate the frequencies, which avoids the complex matching process. The paper is organized as follows: Section 2 gives our analysis and method. Simulation results are shown in Section 3. The last section draws conclusions.
\section{Proposed Method }
\subsection{Problem Formulation}
Consider a complex signal $x(t)$ containing $K$ frequency components with unknown constant amplitudes and phases, and additive noise that is assumed to be a zero-mean stationary complex white Gaussian random process. The samples of the signal at the sampling rate $F_S$ can be written as
\begin{equation}\label{eq1}
 x(n) =\sum\limits_{k = 1}^K {{ s_k}{e^{j(2\pi {f_k}n/F_S)}} + w(n)},n=1,2,\cdots,
\end{equation}
where $f_k$ is the $k$-th frequency, $s_k$ is the corresponding complex amplitude, and $ w(n)$ is additive Gaussian noise with variance $\sigma^2$. Assume that the upper limit of the frequencies is known, but we only have low-rate analog-to-digital converters whose sampling rates are far lower than the Nyquist rate. Undersampling leads to spectral aliasing and frequency ambiguity. Many articles use multi-channel measurement systems to solve the problem. We shall demonstrate that at least three undersampled channels with specific rates can guarantee the success of resolving frequency ambiguity.
\subsection{Unfolding in The Frequency Domain}
Suppose the highest frequency contained in the signal is lower than $f_{H}$, we sample at the rate $F_{S1}=f_{H}/a~(a>1)$, where $a$ is known as the undersampling ratio.\footnote{Actually the Nyquist rate in real number field is $2f_H$, in this paper we assume that complex signals can be sampled directly.} For ease of analysis, $a$ is restricted to be an integer. The collected samples can be written as
\begin{equation}\label{eq2}
x(n) = \sum\limits_{k = 1}^K {{s_k}{e^{j2\pi {f_k} \cdot na/{f_H}}} + w(n)} ,n = 1,2, \cdots.
\end{equation}
If we regard these samples as normal data sampled at the Nyquist rate and process them by methods such as DFT or conventional subspace techniques, a formal estimation of $\hat{f}_k$ will be obtained. If $a=1$, the normalized frequency $\hat{f}_k/f_H$ is the correct estimate. In the case of undersampling, the normalized frequency $\hat{f}_k/f_H$ is actually the estimate of $a\cdot f_k/f_H$, but they can not be one-to-one correspondence because the values of $a\cdot f_k/f_H$ may be greater than 1. In other words, we can not get a unique estimate of $f_k$ from $\hat{f}_k$. Due to the periodicity of trigonometric functions, the estimated normalized frequency ${{a{\hat f}_k}}/{{f_H}}$ differs from ${{a{f_k}} /{{f_H}}}$ by an integer $\kappa$, i.e.,
\begin{equation}\label{eq3}
  {\hat f_k} = {f_k} - \kappa  \cdot {f_H}/a,\kappa  \in \mathbb{N}.
\end{equation}
Without loss of generality, assume that $\hat f_k$ is the minimum value that satisfies (\ref{eq3}) in the interval $(0,f_H)$, all possible eligible frequencies can be unfolded as
\begin{equation}\label{eq4}
  {\tilde f_k} = {{\hat f}_k} + \alpha_{k} \cdot {f_H}/a,~\alpha_{k} = 0,1, \cdots ,a - 1.
\end{equation}
Thus we obtain a series of eligible frequencies from one sub-Nyquist channel. If the true value of $\alpha_{k}$ is solved, the correct estimate of $f_k$ can be found from $\tilde f_k$. Obviously, it's almost impossible to determine the correct frequencies through only one sub-Nyquist sample sequence.
\subsection{The Match of The Frequencies}
In order to resolve frequency ambiguity caused by undersampling, another channel sampled at the rate $F_{S2}=f_{H}/b~(b>1)$ is required. Consequently, another set of the eligible frequencies can be obtained, namely
\begin{equation}\label{eq5}
  {\tilde f'_k} = {{\hat f'}_k} + \beta_{k} \cdot {f_H}/b,~\beta_{k} = 0,1, \cdots ,b - 1,
\end{equation}
where $(\ast)'$ denotes the parameters related to the second channel. For each $k$, at least one value of $\tilde f'_k$ is the same with some value of $\tilde f_k$. In other words, the set composed of ${\tilde f_k}$ and that composed of ${\tilde f'_k}$ contain the same frequency, which is the correct estimate. However, the matchup of the eligible frequencies among different $k$ is unknown. For every $k$, if ${\tilde f_k}$ and ${\tilde f'_k}$ are matched one to one, the correct frequencies will be found.

\begin{figure}[!htbp]
\centering
\subfigure[$f_k$={22Hz, 50Hz}]{
\includegraphics[scale=0.6]{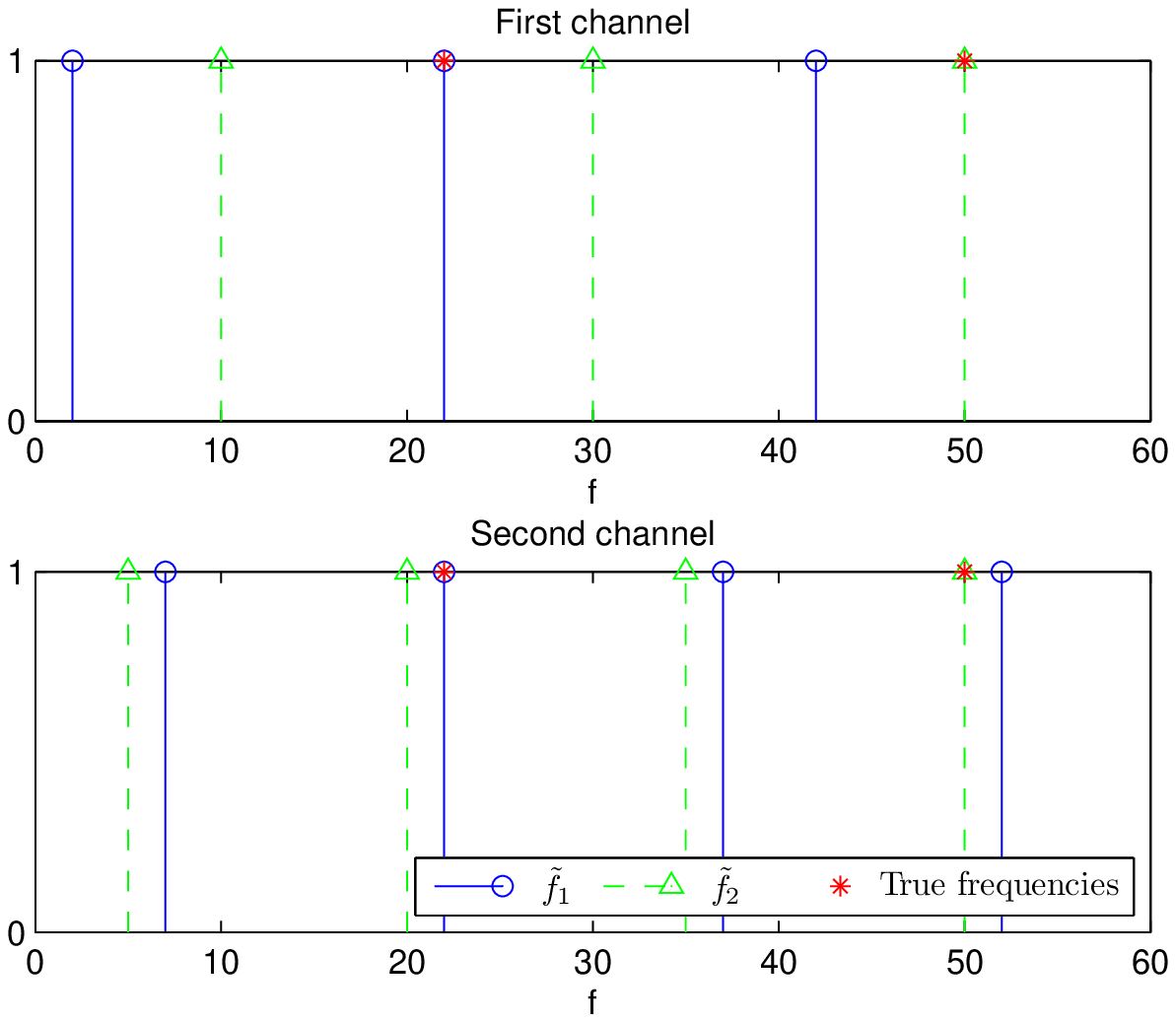}}
\subfigure[$f_k$={25Hz, 50Hz}]{
\includegraphics[scale=0.6]{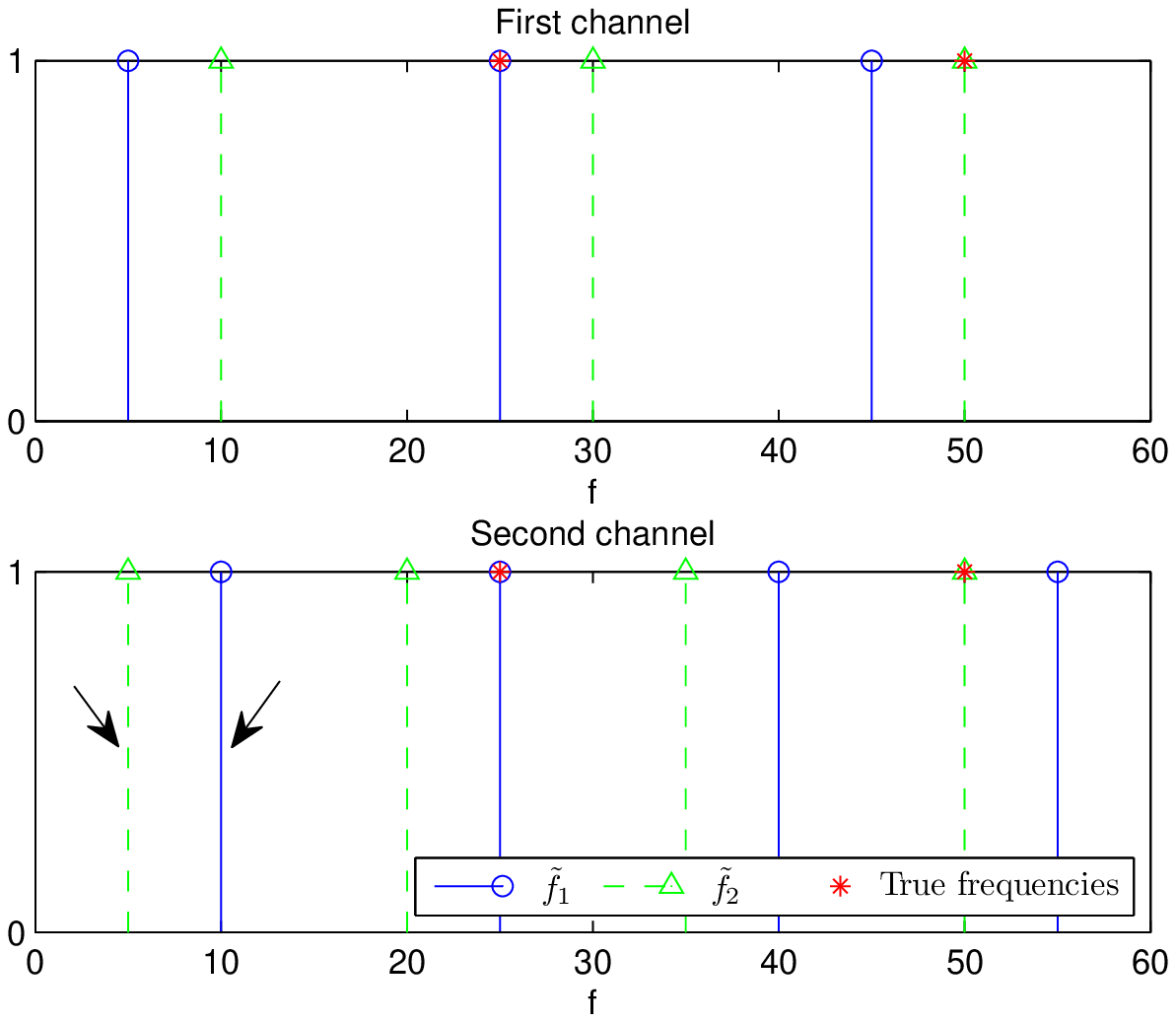}}
\caption{The match of the eligible frequencies obtained from two channels. (a) $f_k$={22Hz, 50Hz}; (b) $f_k$={25Hz, 50Hz}. } \label{fig1}
\end{figure}

To illustrate this process more clearly, we give an example. We assume that the highest frequency in the signal is lower than 60Hz and the undersampling ratios of the two channels are $a=3$ and $b=4$, respectively. The matching process of the eligible frequencies obtained from the two channels are shown in Fig.~\ref{fig1}. In Fig.~\ref{fig1}(a), the true frequencies are taken as 22Hz and 25Hz. Through the first channel, each frequency is unfolded into 3 possible frequencies according to (\ref{eq4}). Similarly, 4 possible frequencies are obtained for each true frequency through the second channel. We need to find the equal values in the eligible frequencies of the same frequency component in different channels. We can see that the two sets of eligible frequencies coincide at 22Hz and 50Hz, which is the true frequencies. However, such a matching process is not always smooth. In Fig.~\ref{fig1}(b), the true frequencies are 25Hz and 50Hz. The two sets of eligible frequencies coincide not only at 25Hz and 50Hz but also at 5Hz and 10Hz. Obviously, matching the eligible values of $f_1$ with those of $f_2$ between the two channels results in an erroneous match. In fact, we can not tell which of the matching results is correct unless we know the true frequencies. This matching process also makes sense for more frequency components.

Next we analyze the matching process of the frequencies. Let
\begin{equation}\label{eq6}
   {\tilde f_m}={\tilde f'_l},~l,m=1,2,\cdots,K,
\end{equation}
we have
\begin{equation}\label{eq7}
   {{\hat f}_m} + {\alpha_m} \cdot {f_H} /a = {{\hat f'}_l}  + {{\beta}_l} \cdot {f_H} /b,
\end{equation}
i.e.,
\begin{equation}\label{eq8}
  b{\alpha _m} - a{\beta _l} = {{ab\left( {{{\hat f'}_l} - {{\hat f}_m}} \right)} \mathord{\left/
 {\vphantom {{ab\left( {{{\hat f'}_l} - {{\hat f}_m}} \right)} {{f_H}}}} \right.
 \kern-\nulldelimiterspace} {{f_H}}}.
\end{equation}
The matching process amounts to solving $\alpha_m$ and $\beta_l$ from (\ref{eq8}). Denoting the true values of $\alpha_{m},{\beta_l}$ by $\bar{\alpha}_{m},\bar{\beta}_l$, we have
\begin{equation}\label{eq9}
   f_{m}={{{\hat f}_m} + \bar{\alpha}_{m} \cdot {f_H}}/a,~f_{l}={{{\hat f'}_l} + \bar{\beta}_{l} \cdot {f_H}}/b.
\end{equation}
Substituting (\ref{eq9}) into (\ref{eq8}) yields
\begin{equation}\label{eq10}
\begin{split}
 b{\alpha_m} - a{\beta_l} =ab\left( {{f_l} - {f_m}} \right)/{f_H} + b{{\bar \alpha}_m} - a{{\bar \beta}_l}.
\end{split}
\end{equation}
To solve the binary indefinite equation (\ref{eq10}), we introduce the following B\'{e}zout's identity \cite{tignol2001galois}:
\newtheorem{thm1}{Theorem}
\begin{thm1}
Let $a$ and $b$ be positive integers with greatest common divisor equal to $d$. Then there are integers $u$ and $v$ such that $au + bv = d$. In addition, the greatest common divisor $d$ is the smallest positive integer that can be written as $au + bv$, and every integer of the form $au + bv$ is a multiple of the greatest common divisor $d$.
\end{thm1}

We focus on the situation that $a$ and $b$ are coprime. According to B\'{e}zout's identity, when $a\perp b$, (\ref{eq10}) has integer solutions as long as its right hand side is an integer. Moreover, since $\alpha_m\in \{0,1,\cdots,a-1\}$ and ${\beta}_{l}\in \{0,1,\cdots,b-1\}$, the equation (\ref{eq10}) just has a unique satisfactory solution. When $l=m$, the unique true values ${\bar \alpha}_m$ and ${\bar \beta}_l$ can be solved. This means that the match is successful and the result of the match is correct. If $l\neq m$ and the right side of (\ref{eq10}) is an integer, (\ref{eq10}) also has a unique solution. However, in this case, the unique solution of (\ref{eq10}) is not the true values $\bar{\alpha}_{m},\bar{\beta}_l$. The match seems somehow plausible but leads to the wrong outcome. This corresponds to the case in Fig.~\ref{fig1}(b). In other words, to make the match correct, the right side of (\ref{eq10}) should be an integer when and only when $l=m$. We find that when $|f_l-f_m|~(l\neq m)$ is an integer multiple of $f_H/\left(ab\right)$, the right hand side of (\ref{eq10}) must be an integer, which may interfere with the correct matching process. Therefore, when two true frequencies $f_l,f_m$ satisfy that $\Delta f\triangleq |f_l-f_m|~(l\neq m)$ is an integer multiple of $f_H/\left(ab\right)$, the path of the match is not unique and the matching result may suffer from mistakes. In fact, when $a$ and $b$ are not coprime integers, frequency matching is almost impossible to complete correctly.

\begin{figure}[!htbp]
\centering
\includegraphics[scale=0.6]{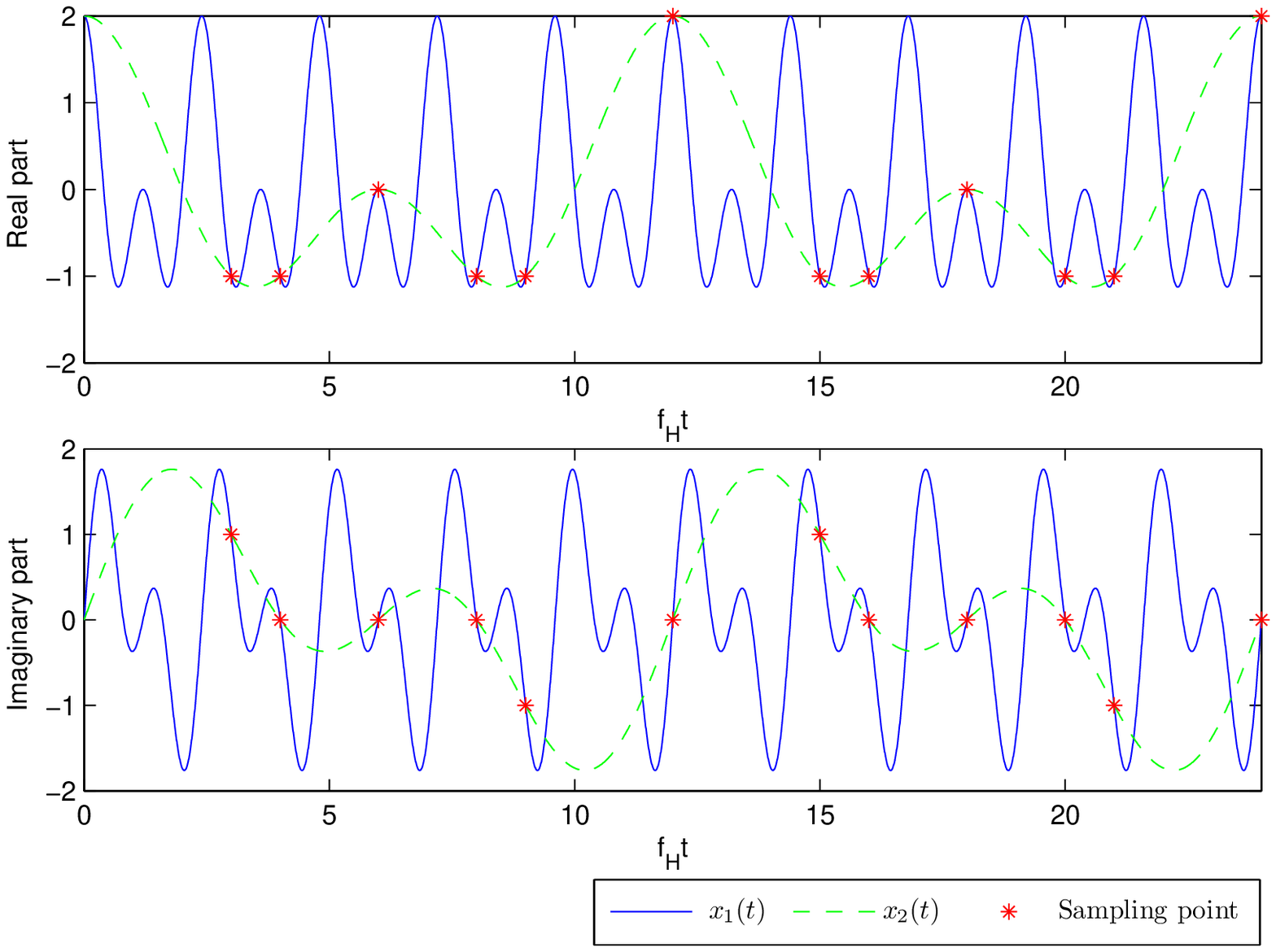}
\centering
\caption{ The signal $x_1$, $x_2$ and the sampling points. } \label{fig2}
\end{figure}

Here we give an example to prove more directly that two coprime channels in some cases can not guarantee the resolution of ambiguity. We construct two complex signals with two frequency components, one is $x_1(t)=e^{j2\pi\cdot25 t}+e^{j2\pi\cdot50 t}$, the other is $x_2(t)=e^{j2\pi\cdot5 t}+e^{j2\pi\cdot10 t}$. We only know that the highest frequency in the signal is lower than 60Hz and we sample at two coprime undersampling ratios $a=3$ and $b=4$, i.e., $F_{S1}=20$Hz, $F_{S2}=15$Hz. As shown in Fig.~\ref{fig2}, we plot (the real and imaginary parts of) the two signals in the time domain and mark the locations of the sampling points. We find that the sampling points of the two signals are identical! That is, it is impossible to uniquely determine the frequencies of the signal by virtue of the samples collected by the two channels.

Next we provide a sufficient condition for deterministic undersampling in the multi-channel sampling framework. Suppose that we have three sub-Nyqiust channels which sample at $f_H/a,f_H/b$ and $f_H/c$ respectively and $a,b,c$ are pairwise coprime. Before the next step we first prove a theorem.
\newtheorem{thm2}[thm1]{Theorem}
\begin{thm2}
Let $a,b,c$ pairwise coprime and $0<f<1$, then there exist no such $f$ that satisfies $f=p/ab$ and $f=q/ac$ and $f=r/bc$ simultaneously, where $p,q,r$ are arbitrary integers.\label{thm2}
\end{thm2}
\newproof{pf}{Proof}
\begin{pf}
From $f=p/ab=q/ac=r/bc$, we obtain $abcf=cp=bq=ar$. This means that $abcf$ is an integer multiple of $a, b, c$ simultaneously. Since $a, b, c$ are pairwise coprime, their least common multiple is $abc$. Therefore, $f$ must be an integer. That is, $f$ satisfying the condition within the range $(0,1)$ does not exist.
\end{pf}

If any two of the three channels are selected to estimate the frequencies of the signal, the match of the eligible frequencies may fail. According to the above analysis, the condition that leads to failure is that $\Delta f=pf_H/ab$ or $\Delta f=qf_H/ac$ or $\Delta f=rf_H/bc$, where $\Delta f\triangleq |f_l-f_m|~(l\neq m)$ and $p,q,r\in \mathbb{Z}$. Consider that the three sub-Nyquist channels are jointly used to estimate the frequencies. Since $0<\Delta f<f_H$, according to Theorem \ref{thm2}, there exist no such $f_l,f_m$ that satisfy $\Delta f=pf_H/ab$ and $\Delta f=qf_H/ac$ and $\Delta f=rf_H/bc$ simultaneously. Therefore, if the three sub-Nyquist channels are jointly used, there must be a match between two channels that is unique, thus the matching result is uniquely determined. That is to say, three sub-Nyquist channels with pairwise coprime undersampling ratios are enough to guarantee the success of the match. It should be pointed out that the problem concerned in this paper is the frequency estimation in the time domain, and the conclusion can not be simply extended to the direction-of-arrival (DOA) estimation in the spatial domain. In the DOA estimation, as described in \cite{Vaidyanathan2011Sparse}, two sets of coprime arrays can greatly increase the degrees of freedom.
\subsection{The Algorithm for Estimation }
In the above subsection, we analyze the matching process between different channels. When the signals contain multiple frequencies, the direct match is intricate and impractical. We fulfill the match by the MUSIC algorithm.

According to the analysis about frequency matching, three channels are required to be used jointly to resolve ambiguity. Assume that three channels begin to sample at the same time point $t=0$, the samples of the first channel are sampled at the time points $a\Delta t,2a\Delta t,\cdots$, where $\Delta t=1/f_H$. Denote the indices of the samples sampled at $a\Delta t,2a\Delta t,\cdots$ as $a,2a,\cdots$, similarly, the indices of all three-channel samples are
\begin{equation}\label{eq11}
 \mathcal{T }= \left\{ {a,2a, \cdots } \right\} \cup \left\{ {b,2b, \cdots } \right\} \cup \left\{ {c,2c, \cdots } \right\}.
\end{equation}
We construct the first snapshot or measurement vector using the samples with indices no greater than $abc$. Let $x[i]$ denote the sample with index $i$, the $l$-th snapshot is formed as
\begin{multline}\label{eq12}
{{\bm {x}}_l} =  \Big(x\left[ {a + abc\left( {l - 1} \right)} \right],x\left[ {b + abc\left( {l - 1} \right)} \right],x\left[ {c + abc\left( {l - 1}\right)}\right],\\
x\left[ {2a + abc\left( {l - 1} \right)} \right],x\left[ {2b + abc\left( {l - 1} \right)} \right],x\left[ {2c + abc\left( {l - 1} \right)} \right],\cdots,\\
x\left[ {abcl} \right]\Big)^T ,~l=1,2,\cdots,L,
\end{multline}
where $(\ast)^{T}$ denotes the transpose operation. Let $\omega_k=2\pi{f}_k/f_H$, ${\bm x}_l$ can be expressed as
\begin{equation}\label{eq13}
  {\bm x_l} = \bm A\bm s_l + \bm w_l,
\end{equation}
where
\begin{equation}\label{eq14}
  \bm A = \left( {\bm a({f_1}), \cdots ,\bm a({f_K})} \right),
\end{equation}
\begin{equation}\label{eq15}
  \bm a({f_k}) = {\left( {{e^{j{\omega _k}a}},{e^{j{\omega _k}b}},{e^{j{\omega _k}c}}, \cdots ,{e^{j{\omega _k}abc}}} \right)^T},
\end{equation}
\begin{equation}\label{eq16}
  \bm s_l = {\left( {{s_1}{e^{j{\omega _1}abc\left( {l - 1} \right)}}, \cdots ,{s_K}{e^{j{\omega _K}abc\left( {l - 1} \right)}}} \right)^T},
\end{equation}
\begin{equation}\label{eq17}
  \bm w_{l} = \Big( w\left[ {a + abc\left( {l - 1} \right)} \right],w\left[ {b + abc\left( {l - 1} \right)} \right],
  w\left[ {c + abc\left( {l - 1} \right)} \right], \cdots ,w\left[ {abcl} \right] \Big)^T.
\end{equation}

The following procedure is the same with conventional MUSIC. The eigenvalue decomposition of the autocorrelation matrix of the data matrix $\bm X = \left({\bm x_1}, \cdots ,{\bm x_L}\right)$ is performed, and then the pseudo-spectrum is obtained by searching on the frequency axis. According to the pseudo-spectrum, the correct estimated frequencies are found. The accuracy of the estimation depends on the search step size. The number of frequencies that can be estimated depends on the matrix $\bm A$, and it can be improved by the method in \cite{Vaidyanathan2011Sparse}.
\section{Simulation Results }
In this section, we verify the conclusions of this paper through experiments. The created signal is a mixture of $K$ complex sinusoids buried in zero-mean complex Gaussian noise at SNR=10dB. We set $K=3$ frequency components with amplitudes of 0.6, 0.7 and 0.8 and random phases. For the sake of brevity, $a=3,b=4,c=5$ and $f_H=60$Hz are set. We use MUSIC for three channels jointly and compare the results with using MUSIC for only two channels. The three channels that sample at $f_H/a,f_H/b,f_H/c$ are called C1, C2, C3 respectively for convenience. In the case of using MUSIC for C1, C2 jointly, the $l$-th snapshot is formed as
\begin{equation}\label{eq18}
{{\bf{x}}_l} = {\Big( {x\left[ {a + ab\left( {l - 1} \right)} \right],x\left[ {b + ab\left( {l - 1} \right)} \right], \cdots ,x\left[ {abl} \right]} \Big)^T},~l = 1,2, \cdots ,L.
\end{equation}
The configuration of samples in the cases of using MUSIC for C2, C3 and C1, C3 can be deduced by analogy. The number of snapshots is $L=100$ and the search step size is 1Hz for all cases.

\begin{figure}[!htbp]
\centering
\subfigure[$f_k$={5Hz, 10Hz, 15Hz}]{
\includegraphics[scale=0.4]{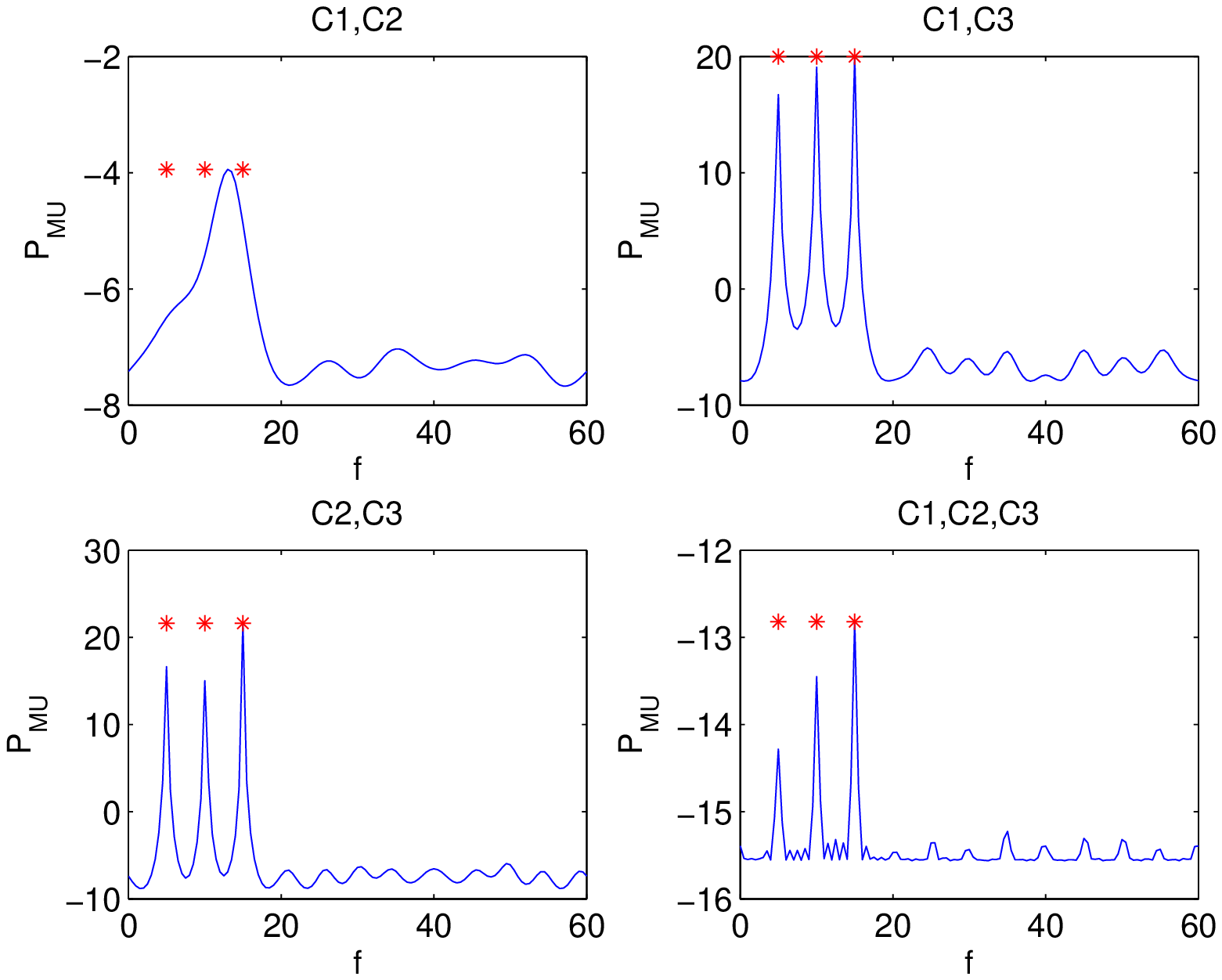}}\\
\subfigure[$f_k$={5Hz, 10Hz, 18Hz}]{
\includegraphics[scale=0.4]{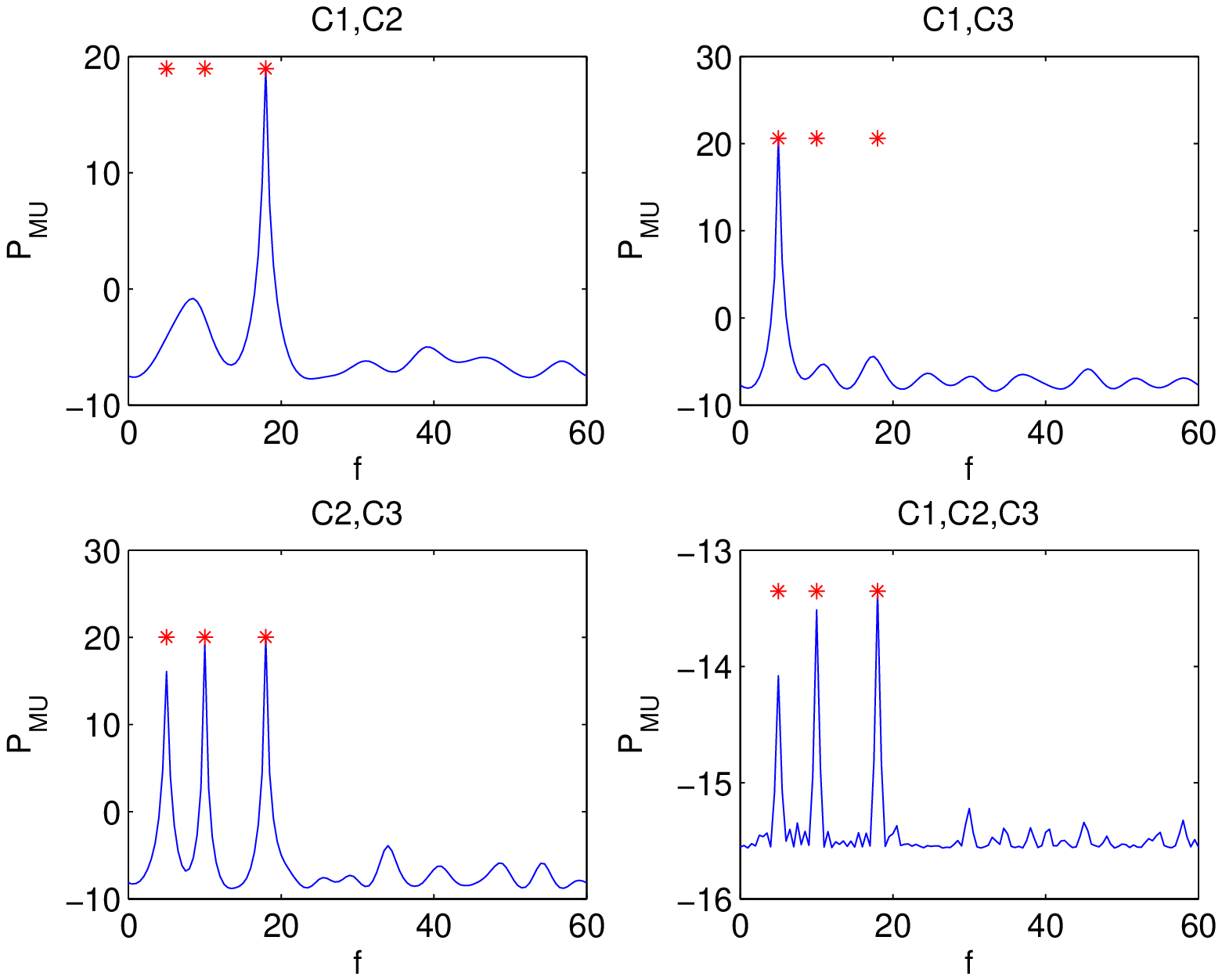}}\\
\subfigure[$f_k$={5Hz, 10Hz, 26Hz}]{
\includegraphics[scale=0.4]{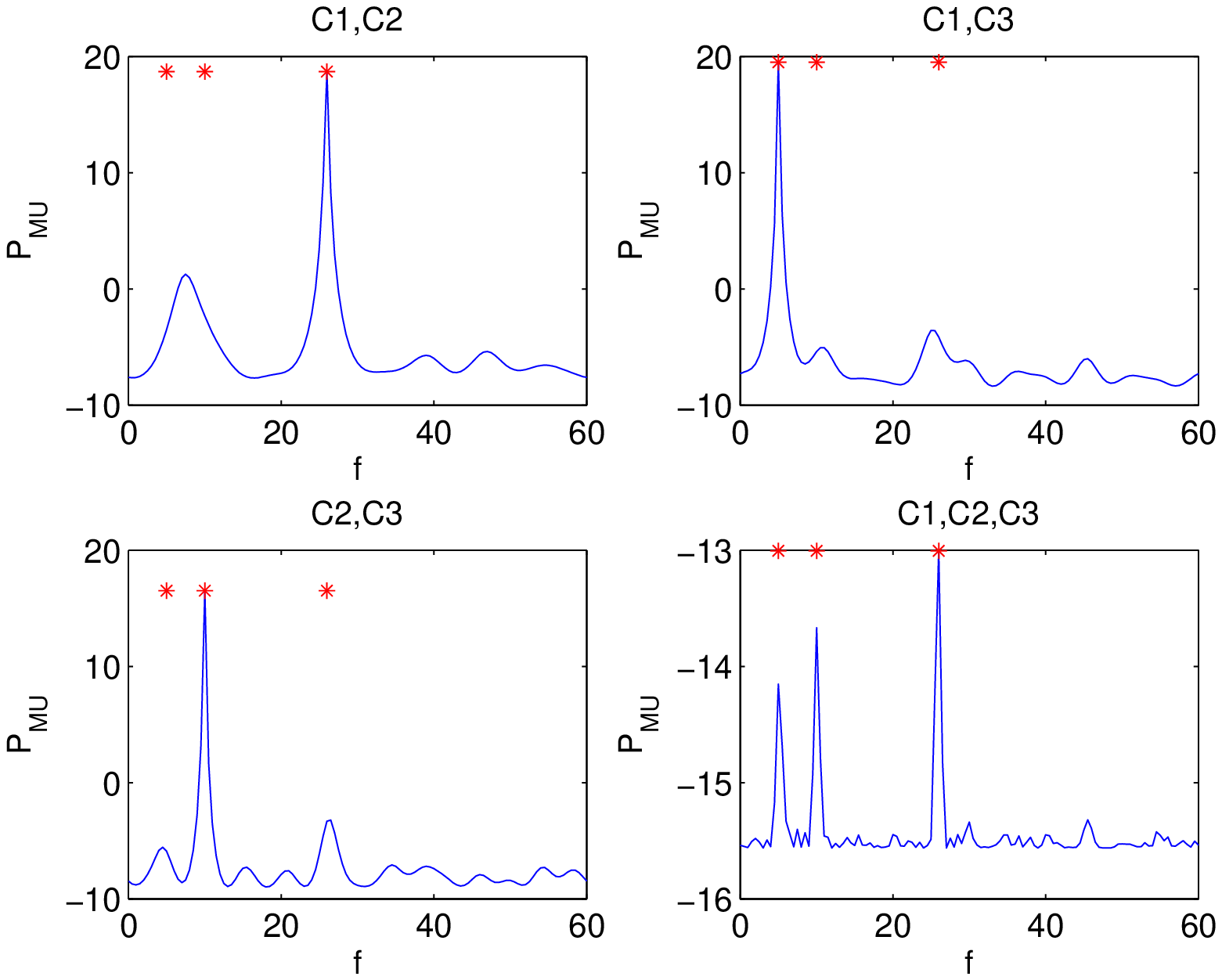}}
\caption{The results of MUSIC for different combinations of three channels. (a) $f_k$={5Hz, 10Hz, 15Hz}; (b) $f_k$={5Hz, 10Hz, 18Hz}; (c) $f_k$={5Hz, 10Hz, 26Hz}. } \label{fig3}
\end{figure}

We set three groups of true frequencies: (a)5Hz, 10Hz, 15Hz; (b)5Hz, 10Hz, 18Hz; (c)5Hz, 10Hz, 26Hz. The results of estimation are shown in Fig.~\ref{fig3}, the red asterisks mark the positions of true frequencies. In Fig.~\ref{fig3}(a), all the intervals of true frequencies are integer multiples of 5Hz. According to our analysis, if one of the intervals of true frequencies is an integer multiple of ${f_H}/\left( {ab} \right) = 5$Hz, using MUSIC for C1, C2 may not identify the true frequencies. As expected, in addition to using MUSIC for C1, C2, the other contrasts identify the true frequencies. In Fig.~\ref{fig3}(b), the intervals of the true frequencies contain a multiple of 5Hz and a multiple of 4Hz, so using MUSIC for C1, C2 and C1, C3 fail. In Fig.~\ref{fig3}(c), the intervals of the true frequencies contain a multiple of 3Hz, a multiple of 4Hz and a multiple of 5Hz. The cases of using two channels all fail and only using three channels jointly yields the correct result. This experiment effectively proves our conclusion.
\section{Conclusion}
In this paper, we have analyzed the choice of the number of channels in multi-channel undersampling with coprime ratios. We demonstrate that three sub-Nyquist channels with pairwise coprime undersampling ratios are generally enough for ambiguity resolution. Then direct MUSIC for three channels jointly is used to estimate the frequencies. The experimental results are in agreement with the theoretical analysis.

\section*{References}
\bibliography{bibfile}

\end{document}